\documentclass[a4paper]{article}
\usepackage{graphicx} 
\usepackage{amsmath}
\usepackage{natbib}
\usepackage{url}
\usepackage{hyperref}

\providecommand{\keywords}[1]
{
  \small	
  \textbf{\textit{Keywords---}} #1
}

\title{Multimodal photometry and spectroscopy: a new approach to data analysis in Astrophysics}

\author{
Johanna Casado,$^{1}$$^{2}$\thanks{E-mail: johanna.casado@um.edu.ar)}
and  B. García$^{1}$$^{3}$
\\
$^{1}$ITeDA (CNEA, CONICET, UNSAM)\\
$^{2}$IBio-University of Mendoza, Mendoza, Argentina\\
$^{3}$Universidad Tecnológica Naciona, \\Facultad Regional Mendoza, Mendoza, Argentina
}
\date{April 2024}

\begin{document}

\maketitle

\begin{abstract}
In the last decade, the multi-sensory approach to data analysis has gained relevance. The possibility of including people with vision difficulties in the field of education and the dissemination of science is part of it. However, in the field of scientific research, the topic is not yet accepted, mainly due to the lack of conclusive evidence related to its robustness and performance, except for a few exceptions.
On the other hand, very few of the tools that allow data to be sonifyed are focused on the user or have an interface that allows controlling the sound production process and, at the same time the visualization of data. Of those that meet this requirement and present application in the field of astronomy and astrophysics, only four have included people with visual disabilities in the design and only two of them have carried out tests with users in Focus Groups during its development.
This work details the characteristics of sonoUno, with its end-user-focused design, its capabilities in specific use cases in astrophysical data analysis related to spectroscopy and photometry, following the basic software functionalities while also introducing some innovation.
\end{abstract}




\section{Introduction}

Human beings by nature explore the world through all their senses, however, is predominant the use of visualizations to make sense of the data sets under study. This is the case, even in astronomy, where most of the data are outside the visible range, and, evidence is found about the benefits of auditory display as a complement to visual display \citep{wandathesis:2013}.

The use of sound/audio in astronomy has existed for years; some 
examples are:  the zCOSMOS astronomical data set sonification, where the authors describe the data set and the sonification strategy used \citep{bardellietal2021}; LightSound, an electronic device that converts light into sound, created to observe eclipses \citep{lightsound2020}; a sound platform built in collaboration with the ATLAS Outreach team and a website that allows the general public to listen to experiments in real time \citep{cherstonetal2016}; the Quasar Spectroscopy Sound software focused on sonify cosmological data sets, the technique used is described by \citet{hansenetal2020}; a project with the refunctionalization of two cosmic particle accelerators where the use of auditory display in conjunction with visual display was described, the objective was to make the discoveries more accessible, the sound was made with musical characteristics and developed specifically for this case study \citep{ohmetal2021}; \citet{quintonetal2021} describes a sonification design for planetary orbits in asteroid belts. In the majority of cases, the mapping carried out to obtain the sonification of the data set is defined by its creator and shared as a final product.

Between all the different projects, groups, and sonification techniques, some works by visually impaired people arise. \citet{solarwindWanda} describes a study of the EX Hya light-curve and Solar Wind using xSonify. In the article, the software is described and also explains how to interpret the data using sonification. Moreover, \citet{starsound2022} manifests how he continues working as a researcher using the software StarSound and VoxMagellan. They explain the use of StarSound to sonify high-redshift galaxy data and the Deeper, Wider, Faster (DWF) program. For more complex data, for example, a scatter plot that contains galaxies dispersed in color (one axis) and magnitude (another axis), they used VoxMagellan with a parameter sonification that represents the different parameters. These examples reinforce that sonification could be used to research and generate inclusion in the field; however, it is a challenge for the technique to be universally adopted.

Devoted to sonification in general, since its creation in 1992, the purpose of the International Conference on Auditory Display (ICAD) has been to bring together multidisciplinary experts working in the field of sonification. This conference presents a repository where there is a large number of works that \citet{andreopoulouYgoudarzi2021} grouped into six categories: sonification methods, sonification tools/systems, review/opinion studies, exploratory studies, perception/evaluation studies, and others. This systematic review of the ICAD repository highlights a high percentage of articles dedicated to sonification methods and tools, in contrast to a low percentage concerned with design methodologies, perception studies, evaluation methods, and astronomical data analysis using this technique. It is alarming that perception studies showed growth between 2005-2009, but decreased below 1\% by 2019, even when \citet{fergusonYbrewster2017} pointed out the importance of perception studies in auditory displays and reported some psychoacoustic parameters mentioning how people perceive the sound. These works were complemented recently by \citet{perCS2023}, where the study of the full text of the ICAD proceeding from 2017 to 2022 was done. Given the fact that the keyword perception presents a low rate of use: 6 times less than `auditory', 3 times less than `display', and 2 times less than `music'.

The novelty of the technique and the low case of use inside the research file in conjunction with visualization to validate it, raise concerns about whether the technique may be biased. \citet{supper2014} describes the difference between visual and audio displays and how both are perceived in terms of the immersive and emotional environment. Along the same lines, \citet{neuhoff2019} raises the challenge of sound versus visualization at a perceptual level, pointing out the high percentage of works that use the word ``music'' when talking about sound for data deployment (remember the rates of \citet{perCS2023}). These concerns, added to the lack of studies that illuminate the understanding of how the brain interprets data sonification, complicate the fact that the sonification technique is adopted as an analysis tool in the scientific field. This reinforces the need for robust sound developments and perception studies to learn how people understand acoustic parameters, as well as training to learn to listen and understand scientific data sets, in order to detect features and signals in such data using sonification.

Remarkably, a work by \citet{MNRAStucker2022} uses the software Astronify (see section \ref{sec:soniftools}) to study light curves. They describe the sonification process, how some sample data was made, and the process of testing with participants. The participants had no experience with sonification, and the results showed that experts perform better with plots; on the other hand, experts and non-experts present no difference using sonification. The last may reinforce the lack of training in this technique that both groups present. Now using STRAUSS (see section \ref{sec:soniftools}), last year \citet{trayford2023} presented the use of spectra sonification to evaluate if participants could rate A) variation in SNR, B) variation in emission-line width, and C) variation in emission-line flux ratio. Under 58 respondents, the ratings present a relevant correlation with the physical properties presented. The authors express that, given the minimal training and small sample, these are very promising results.

The framework described and the last investigations focused on using sonification and testing its efficiency, emphasize the importance and needs of investigations centred on sonification as a technique that could complement the current display of astronomical data. In addition, it is important to have a tool that allows this technique to be carried out autonomously by people with and without disabilities.

This work resumes the sound model of the sonoUno software, which aims to provide an open-source platform that allows users to open different sets of data and explore them through a visual and auditory display \citep{casadoJOSS2024}. Examples of application with two Galaxies are shown, including the possibility of image sonification, spectra sonification, special features detection, and comparison of these two spectra through sonification. In addition, using variable stars, the plots of the database and sonoUno were compared and contrasted with their sounds. Concluding with the possibility of spectral classification using sonification. These examples of application in the field of astronomy aim to show the possibilities of development proposed by research in multi-modal astronomy.

\section {Sonorization tools}
\label{sec:soniftools}

Over the past 10 years, there has been a large increase in the number of projects using sonification to represent astrophysical data. \citet{zanella2022} showed that 98 sonification projects had been developed since 1962; many of them were discontinued, lacked documentation, or had no evidence of applications in science. The majority of sonification projects, almost 80\%, have been carried out between 2011 and 2021. Not all of them share the same objective: some are tools to produce sound through the command line, others have the purpose of offering the user the ability to modify the sound configurations to achieve a sound system that fits their needs, and others prioritize the development of an accessible graphical interface.

One of the tools analyzed in 2017 at the beginning of the development of sonoUno was xSonify. This program was taken as a reference because it presented a development framework focused on the user and inclusion. In addition, the authors had contact with one of its developers. A notable feature was that xSonify allowed the sonification of multiple columns, varying the instrument between each column to be sonifyed. Furthermore, at that time, it was one of the few open-source tools that could be used with a screen reader.

StarSound is a program that began its development at the same time as sonoUno, contact has been maintained with its developers. Initially, they presented this program with a graphical interface that allows the configuration of the sound process in detail, similar to a sound equalizer. An accessibility feature they added was the ability to modify all parameters using a text file for people with visual disabilities. Subsequently, to provide sonification for more complex graphics and images, they created VoxMagelan, a software that allows scatter plots to be imported and provides sound for the region where the pointer is located, allowing the plot to be spatially explored by sonification \citep{starsound2022}.

Concerning STRAUSS, it is a Python package, open source, and available on GitHub\footnote{STRAUSS: \url{https://github.com/james-trayford/strauss}}. It has been developed with the aim of improving the current visual display of data and accessibility. It presents the possibility of being used by people without knowledge of sound or programming, through predefined examples and using the default configurations. However, it offers documentation that allows its use at a more advanced level, allowing modification of the sound parameters. A direct application of this development is in the ``Audio Universe'' project, in which they have recently published a tour of the solar system \citep{strauss2021}.

Some other Python packages devoted to data sonification are Soni-py\footnote{Soni-py: \url{https://github.com/lockepatton/sonipy}} \citep{soni-py2021}, Astronify\footnote{Astronify: \url{https://github.com/spacetelescope/astronify}} and SoniScope\footnote{SoniScope: \url{https://github.com/fhstp/soniscope-jupyter}}. Three of them could be used in a Python environment, a tool widely used to analyze data sets. Soni-py and Astronify were developed and used with light curves, but they are not limited to them. About SoniScope, it sonify regions pointed out with the mouse on a scatter plot \citep{soniscope}. Something remarkable about Astronify is the game they offer with its documentation, it has two levels and presents different characteristics of a light curve; this is the first tool presenting something comparable with a training task.

Now migrating to developments with a graphical web interface exists: Highcharts Sonification Studio\footnote{Highcharts: \url{https://sonification.highcharts.com/}}, Afterglow Access (AgA)\footnote{Afterglow Access: \url{https://idataproject.org/resources/}} and sonoUno\footnote{sonoUno: \url{https://dev.sonouno.org.ar/en-US/}}. The first is a website developed with four objectives as a guide \citep{highchartsonifstudio2017}: (1) maximize accessibility; (2) easy tool for beginners without reducing its high performance; (3) maximize portability; and (4) maximize its utility. \citet{highchartsonifstudio2021} describes the joint work between Highsoft and the Georgia Tech Sonification Lab to develop the tool. Notably, `Highcharts Sonification Studio' is based on almost 20 years of experience in sonification by the Georgia Tech Sonification Lab and the previous development of the Sonification Sandbox software \citep{sandbox2007}. Particularly, Highcharts Sonification Studio has two tabs, one for the data and another that contains all the control elements over the visualization and sound.

AgA, using a different initial approach, presents as its main objective the analysis of astronomical images through sonification. The tool was developed by a multidisciplinary team (high school students, undergraduate students, astronomers, software engineers, and educational researchers, among others) and included people with and without visual disabilities. \citet{afterglow2020} describes the process carried out to achieve a user-centered tool. Some of the techniques used were role models, exercises with questions such as `What?-If' in search of a creative flow of ideas, and direct or indirect contact through GitHub between users, designers, and developers, among others. This marks a clear and published precedent, in the field of astronomy, about the need and importance of user-centered design to create useful, usable, and efficient tools for data sonification.

As for the sonoUno program, which is described and used in this contribution, it is a software proposed, initially, for astronomical and astrophysical data sonification, focused on the user from the beginning. It has a desktop version (currently available) with a graphical interface tested by people with and without disabilities \citep{casadoFG2022} that allows to open tables with two or more columns, display the graph, and allow sonification \citep{casadoJOSS2024}. 

In 2019, the development of a web interface \citep{hcii2022} with similar functionalities to the desktop version began. This web version follows the same principles, its accessibility with screen readers was tested and counted with several user cases. During The Audible Universe 2 workshop, this version was evaluated by experts during a sonification software evaluation task \citep{au2023}, they define sonoUno web as: ``An online tool, or standalone user interface. Suitable for one-dimensional data, such as light curves and spectra. Primarily for scientific analyses and education (add. accessibility focus).''. From there some requirements as a problem with some slide bars arise. Then, during a degree thesis in 2023 \citep{tfBelen2023} a user experience evaluation of the sonoUno web was made. A Focus Group and interviews were conducted with nine participants. The principal result was a new web interface design that is being implemented.

Taking into account the programs presented, very few are focused on the user or have a graphic interface that allows controlling the sound production process and, at the same time, the visualization of data. Of those that meet this requirement and are applicable in the field of astronomy and astrophysics, only four (sonoUno, AgA, StarSound/VoxMagelan, and xSonify) have included people with visual disabilities in the design. Refining the detail a little more, only one tool besides sonoUno (AgA) has exchanged with users during its development, and it only has a web version.

\section{sonoUno visual deployment and sound synthesis}

The sonoUno framework was developed from the beginning to ensure simple access to information for blind and non-blind people. According to that, the deployment of all the tools is made on the same screen, without pop-up windows. The organization of the functionalities (plot, sound, mathematical, and IO functions) are located on different panels, which can be manipulated (hidden or shown) to simplify the display as the user desires. The possibility of using screen readers and shortcut keys improves access; these characteristics are tested after each update.

As the software was conceived as a tool for multimodal analysis of data, it is not only for sonification, and because of that, the graphical output was also important during the development. Indeed, the synchronization of the visual and sound display is very important, a lot of work and development was done to ensure that.

\subsection{Graphical tool}
\label{sec:tool_graph}

Matplotlib\footnote{Matplotlib: \url{https://matplotlib.org/}} is used to produce the plots. This Python library was widely used in astronomy and other science in general. Only as a reference, \citet{matplotlin-python} describes the visual data analysis in their book with Matplotlib. In addition, during the focus group and exchange with expert sonoUno users, most of them expressed that they use this library for data representation. Matplotlib is very versatile, allowing it to produce almost any kind of plot.

The integration of Matplotlib with the graphic user interface (GUI) using wxPython was relatively easy and several examples exist online. The hard part was the synchronization between the position marker on the plot and the reproduced sound for the sonoUno. To solve that, the event function of wxPython is used; this functionality is associated with a timer, something that runs automatically every predefined period. After some tests and taking into account the time of sound reproduction and GUI behavior, the minimum time to call the new event was between 50 ms and 100 ms (it depends on the computer's characteristics too). The consequence was that for a large amount of data, the reproduction time in the GUI was very high; for example, using equation \ref{eq:GUI-rep-time1} for a galaxy spectrum with around 3800 rows and a time between notes of 50 ms, the time is 3,2 minutes (see equation \ref{eq:GUI-rep-time2}, the next \href{https://youtu.be/L1jwbrKiTw8}{video} shows that the time matches).

\begin{equation}
    rep\_time = (\frac{rows * time\_between\_points}{1000})/60 
	\label{eq:GUI-rep-time1}
\end{equation}

\begin{equation}
    rep\_time = (\frac{3840 * 50 ms}{1000})/60 = 3,2 minutes
	\label{eq:GUI-rep-time2}
\end{equation}

The long reproduction time of data sets was criticized a lot by users because it takes a lot of time in only one file during a research process. However, the synchronization between plot and sound to a multimodal display is one of the innovations that sonoUno team wants to maintain; the group is investigating a way to accelerate the process without compromising the synchronization using an open-source language like Python. 

Another big challenge of the development was integrating around six groups of functionalities in the same GUI without pop-up windows. \citet{casadoFG2022} explains the user center design of sonoUno and the theoretical framework used to organize the GUI. First, all the functionalities were grouped by similarity following four casual parameters defined by \citet{IAUS358-casado}. In consequence, the four groups of sonoUno functionalities were: Data display, Data operation, Data configurations, and I/O options. With the groups, four principal panels were programmed with the possibility of showing or hiding them, as needed. The result was a clean first GUI, where only the plot and reproduction options are visible, reducing the memory overload, and better interaction with screen readers. Then, the user could open other panels from the menu bar or shortcut keys. Finally, all the functionalities are presented in the same windows, avoiding losing the keyboard focus by screen reader users.

\subsection{Sonorization}

The sonification process is based on translating certain parameters or information into non-speech audio. This definition also includes the so-called auditory icons and earcons. For this work, only reference will be made to the conversion of information from numerical tables of two or more columns related to astronomy (particularly the spectra of galaxies and variable stars). Until now, it has been taken as a convention to relate the independent variable (which is graphed on the x-axis when performing a visualization) with the sonification time, which means that each data point is translated into sound one after the other, ordered and sequentially. The dependent variable (which is located on the y-axis when performing a visualization) is generally related to frequency, which allows increasing or decreasing changes in the data to be related to the increase or decrease of the sound frequency.

Following the described process, sonoUno generates the waveform of the sound with the equation \ref{eq:soundwave}. By default, and to perform the sonifications in this work, the sine waveform is used to simulate a piano, not limiting the frequencies to 88 keys. In equation \ref{eq:soundwave}, four parameters can be varied, consequently modifying the sound generated.

\begin{equation}
    sound = A \sin{(2ft)}
	\label{eq:soundwave}
\end{equation}

From equation \ref{eq:soundwave}, if \textit{t} is varied, the duration of the sound is modified, what sonoUno calls `tempo'. This variable has a direct relationship with the sonification time, since a greater tempo results in a greater duration of the sound. However, the relationship presents a minimal time between notes for a correct GUI functioning. Given the fact that sonoUno generates one sound per data, that is, point by point, and in each data iteration, it also attends to any event that happens in the graphical interface (for example if the user presses a button), the minimal time per notes is around 50 milliseconds (see section \ref{sec:tool_graph}). The latter indicates that the minimum duration of said tempo will be linked to the processing speed of the computer where it is executed. For this reason, sonoUno in its desktop version has a slow playback speed, a fact that has been highlighted by its users on several occasions \citep{casadoiau367}. However, its developers have highlighted that the main objective is to allow data analysis, which has been shown to be possible \citep{morales2024}.

On the other hand, continuing with equation \ref{eq:soundwave}, varying \textit{f} and \textit{A}, changes in the tone and volume of the sound are achieved respectively. sonoUno allows one to select between these two variables to relate to the y-axis of the data plotted. To start the sonification process, independently of the parameter adopted to sonify the y-axis, `min-max normalization' is used following equation \ref{eq:normalization} (\textit{X} refers to the vector of the tuple to be sonifyed) to bring all the values contained in the dependent column between 0 and 1. This makes it possible to map the information of the celestial object in frequency (multiplying it by \textit{f}) or volume (multiplying it by \textit{A}). The link to some audios will be included in the section \ref{sec:data_analysis} and could be found on the sonoUno gallery\footnote{sonoUno gallery: \url{https://www.sonouno.org.ar/gallery-index/}} too.

\begin{equation}
    X' = \frac{X - X_{min}}{X_{max} - X_{min}}
	\label{eq:normalization}
\end{equation}

\subsection{Feature detection}

The focus of sonoUno is to display in an effective and accessible way any dataset. However, with its use, the need to add special functionalities for data analysis has become evident, and it is part of the permanent improvement of the development. One of these new tools is the detection of peaks in the data.

The first approach to detection was implemented in the sonoUno web interface; the two versions of the software (desktop and web) maintain the same theoretical framework and similar functions but, at the moment, don't share the same programming language \citep{sonounowebHCI2022}. For this, the translation of the `peak finder' function implemented in sonoUno web to Python was needed.

This tool receives the data table and returns an array with the positions where possible emission and/or absorption lines were detected. To search for the peaks, first, a medium line is calculated, taking all the data in the `y' column; then, a sensitivity percentage given by the user is defined to construct a gap from the medium line, up and down; and finally, the coordinates of the maximum or minimum value found above and below this gap are stored.

The determined coordinates of the pics are plotted in the graph together with the data. In addition, a csv file with the x-y coordinates of the peaks could be saved. This function is useful for chemical element identification (see Figures \ref{fig:barred_spiral_spectrum1} and \ref{fig:double_nucleus_spectrum1}).

\subsection{Use of the software in bash}

Responding to a general request from users, a bash script was developed to improve the speed and efficiency of the data-sonification process. The first approach involved searching for all CSV or TXT files in a specific directory (with the location passed as an argument by the user) and then saving the sound and plot (if requested) relating to the first two data columns of each file, in the same folder.

The first approach shows its efficiency in producing the sonification of a lot of data sets, to then perform the analysis. The audio total time is shorter because there is no need to attend the GUI; in addition, this sonification could be accelerated externally too. Nevertheless, not all data types are equal, the general process is the same but there needed some adjustments between the spectrum Galaxy, the image, the phase diagram of a variable star, and spectral star classification sonification. In section \ref{sec:data_analysis} data analysis of each data type and results obtained from sonoUno tools will be shown.

\section{Data analysis and results}
\label{sec:data_analysis}

The sonoUno tool has been adopted to visualize and sonify data of galaxies obtained from the \textit{Sloan Digital Sky Survey} (SDSS)\footnote{SDSS: \url{https://cas.sdss.org/dr18/}}, variable stars obtained from the \textit{All-Sky Automated Survey for Supernovae} (ASAS-SN)\footnote{ASAS-SN: \url{https://asas-sn.osu.edu/variables}}, as well as data of spectroscopy standard stars in order to use the resource to introduce blind people into the spectral classification of stars. The choice of the tool is based on the fact that sonoUno is versatile, open source, developed in Python, and has a user-centered design from the beginning \citep{casadoJOSS2024}. The sonoUno team has carried out user tests evaluating the graphical interface (both desktop and web), managing to adjust it to the needs of users with and without disabilities \citep{casadoFG2022,tfBelen2023}.

\subsection{Galaxies}
\label{sec:galaxies}

The SDSS database has an extensive number of images, spectroscopy in the visible range, and spectroscopy in the infrared, among others. These data have been obtained mainly by the project's telescope located at the Apache Point observatory, New Mexico, United States (``Sloan Foundation 2.5m Telescope''); in the last stages, it has included the ``Irénée du Pont Telescope'' located at the Atacama region in the Andes mountain range, Chile, thus obtaining images of the southern hemisphere.

Regarding accessibility, although the SDSS website has been improved and modified over the years, use with screen readers continues to be an issue for improvement. The interface elements in the SDSS website are better described for screen readers in the last update, however, the discourse presented by NVDA for example continues to lack order and meaning. In some cases, things are presented in reverse and there are still elements that are difficult or impossible to access with screen readers.

As for the data file, when downloading the galaxy spectrum, it contains four columns: the first corresponds to the wavelength, which will be used as the abscissa coordinate (in the case of this data the wavelength range is that of the visible spectrum); the second corresponds to the flux, which represents the amount of light that reaches the instrument at each wavelength; the third is the same flow but some filter or smoothing has been applied to reduce the noise; and the fourth is the light intensity of the sky at each wavelength, it presents background information that is perceived in all observations and includes the atmospheric effect.

\subsubsection{Object SDSS J115537.98-004614.2}
\label{sec:barred-spiral}

\begin{figure}
	\includegraphics[width=\columnwidth]{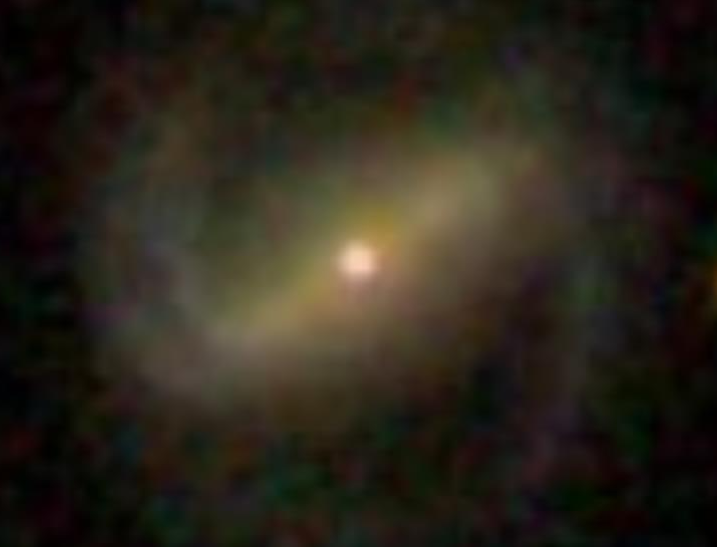}
    \caption{Barred spiral galaxy image extracted from the      \href{https://skyserver.sdss.org/dr17/VisualTools/navi?ra=178.908277828839&dec=-0.77061784547085&scale=0.2}{SDSS database}.}
    \label{fig:barred_spiral}
\end{figure}

The Object SDSS J115537.98-004614.2 is a barred spiral galaxy (More information: \href{https://skyserver.sdss.org/dr17/VisualTools/quickobj?objId=1237674649391333465}{here}). Figure \ref{fig:barred_spiral} shows an image of this object captured from the SDSS database that could be appreciated from the Visual Tool Navigate. Using an image sonification script developed as part of the sonoUno project \citep{IIIWAIcasado2023}, a video with the image sonification was generated and could be found \href{https://youtu.be/3CC2Jl0Vw9U}{here}.

\begin{figure}
    \includegraphics[width=\columnwidth]{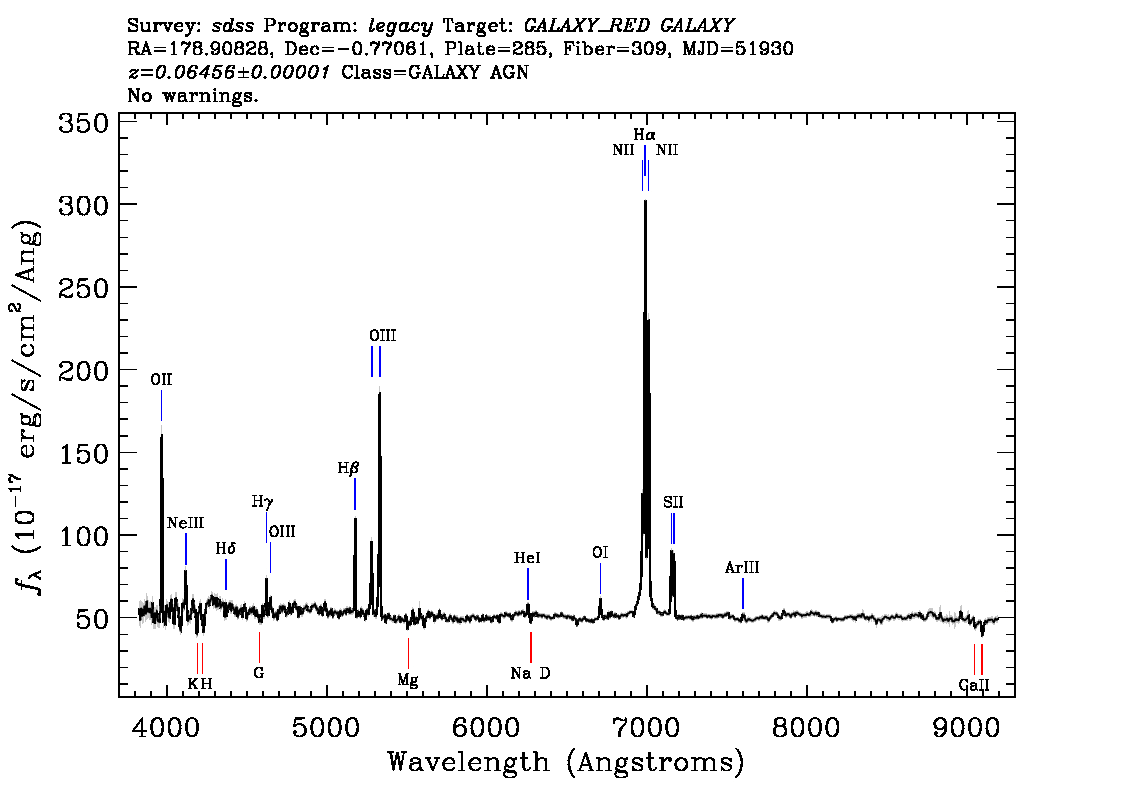}
	\includegraphics[width=\columnwidth]{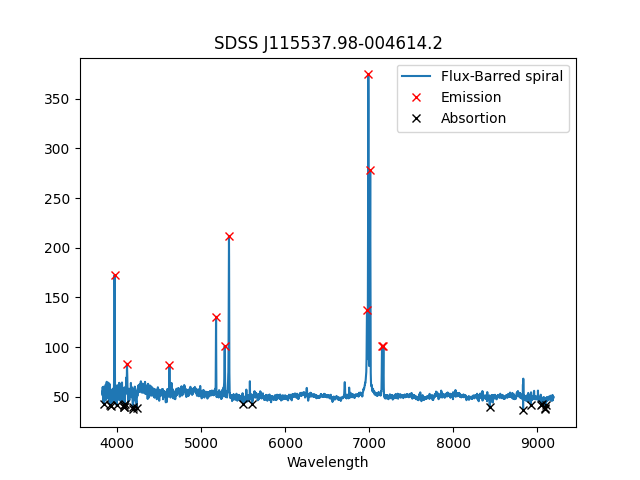}
    \caption{Barred spiral galaxy Flux vs Wavelength plot, top from the database and bottom from sonoUno, with pick detection (\href{https://www.sonouno.org.ar/wp-content/uploads/sites/9/2024/01/SDSS-J115537.98-004614.2_sound-Flux.wav}{link} to the sound).}
    \label{fig:barred_spiral_spectrum1}
\end{figure}

The object page on the database contains a link to download the spectrum as CSV, that file could be opened with sonoUno to display the plot and sonification. To be able to compare the sonoUno display, a combination of the spectrum obtained from the database and that plotted by sonoUno is shown in Figure \ref{fig:barred_spiral_spectrum1}. It could be appreciated that the plots exhibit the same absorption and emission lines.

\subsubsection{Object SDSS J170437.70+603506.4}
\label{sec:double-nucleus}

\begin{figure}
	\includegraphics[width=\columnwidth]{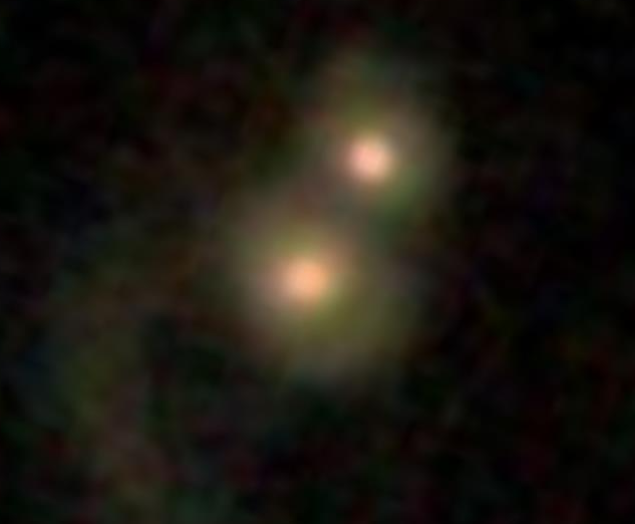}
    \caption{Double nucleus galaxy image extracted from the \href{https://skyserver.sdss.org/dr17/VisualTools/navi?ra=256.157116458609&dec=60.5851296110757&scale=0.2}{SDSS database}.}
    \label{fig:double_nucleus}
\end{figure}

The Object SDSS J170437.70+603506.4 is a double nucleus galaxy (More information: \href{https://skyserver.sdss.org/dr17/VisualTools/quickobj?objId=1237651225171197972}{here}). Figure \ref{fig:double_nucleus} shows an image of this object captured from the SDSS database. Using the same image sonification script mentioned above, a video with the image sonification was generated and could be found \href{https://youtu.be/uDPWlBCzAXk}{here}.

\begin{figure}
    \includegraphics[width=\columnwidth]{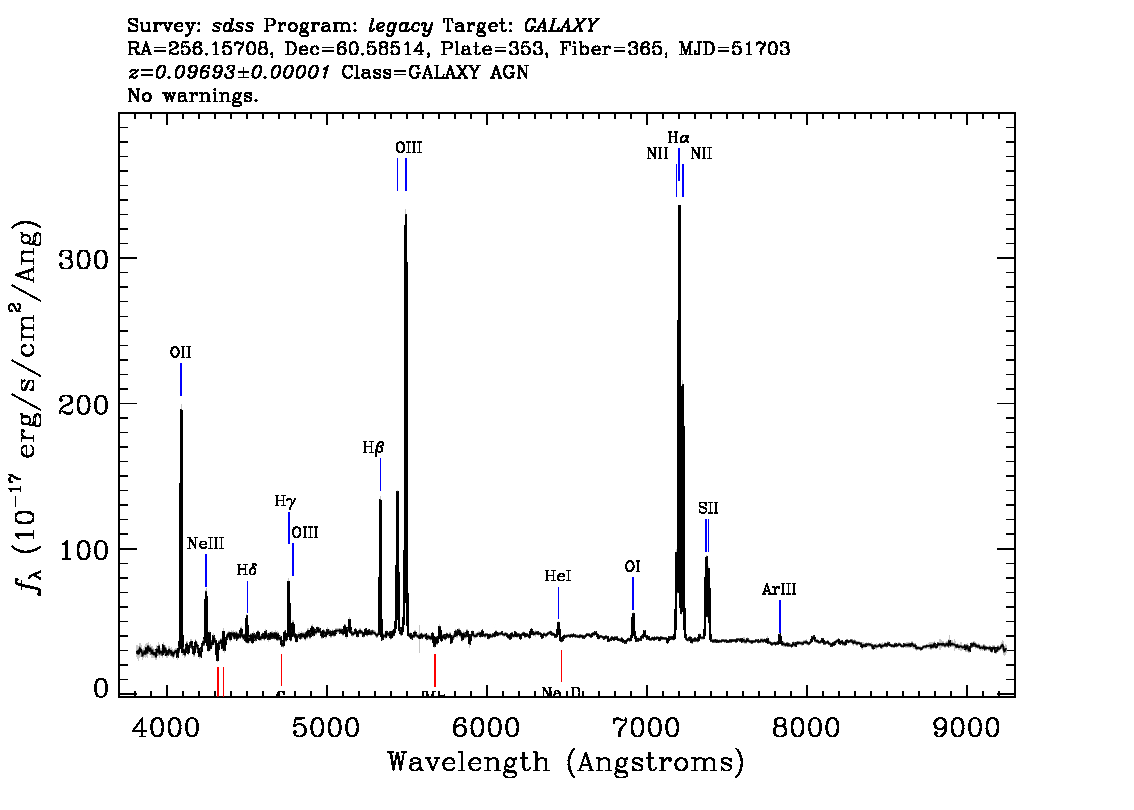}
    \includegraphics[width=\columnwidth]{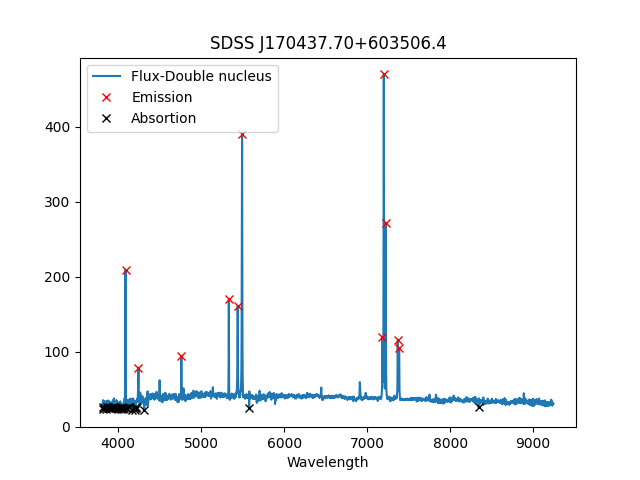}
    \caption{Double nucleus galaxy Flux vs Wavelength plot, top from the database and bottom from sonoUno, with pick detection (\href{https://www.sonouno.org.ar/wp-content/uploads/sites/9/2024/01/SDSS-J170437.70603506.4_sound-Flux.wav}{link} to the sound).}
    \label{fig:double_nucleus_spectrum1}
\end{figure}

Moreover, the CSV downloaded from the database was opened in sonoUno to produce the plot and sonification. Repeatedly, to be able to compare, Figure \ref{fig:double_nucleus_spectrum1} shows the spectrum obtained from the database first, and then the spectrum saved from sonoUno software. Equal to the previous galaxy, the plots exhibit the same absorption and emission lines.

\subsubsection{Comparison of two galaxies spectrum}

A common practice is to compare two or more spectrums between them, something that is not common to do by sonification yet. However, it is known that the human brain can distinguish between a few sounds coming from different sources at the same time, being able to discriminate between information of each one of them. Although from practice we could think that several sounds could be differentiated, it must be remembered that in this case, we are talking about the number of sounds that can be analyzed simultaneously.

SonoUno allows the generation of multicolumn sonification with a particular script, it is not integrated on the GUI yet. SonoUno team opened a GitHub repository with this tool first approach\footnote{SonoUno multicolumn sonification: \url{https://github.com/sonoUnoTeam/sonoUno-multicolumn}}. Some modules of the sonoUno main repository were used, these are import, export, transformation, and sound modules. So the novelty is the use of two different instruments to sonify two curves simultaneously. 

In this contribution, this tool is used to compare the spectra of the two galaxies presented above (section \ref{sec:barred-spiral} and \ref{sec:double-nucleus}), a video with the reproduction could be found \href{https://youtu.be/bhuCK4PuGeE}{here}. The instruments used were 'sine' and 'flute'; one important future work in sonoUno is to augment the offer between instruments in its sound library.

\begin{figure}
	\includegraphics[width=\columnwidth]{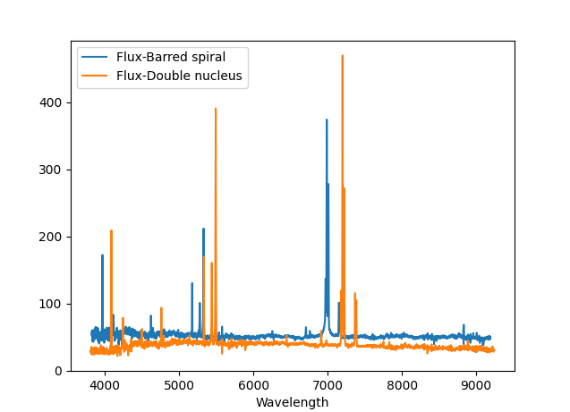}
    \caption{Image obtained from sonoUno script with the spectrum of the two galaxies presented here in the same display.}
    \label{fig:galaxy-spectrum-comparison}
\end{figure}

Although this first approximation was achieved, it has not yet been formally presented within the sonoUno functionalities due to two reasons: (1) to include it in the actual GUI a new design must be made to include the configuration of parameters for each of the graphs, with their consequent user testing; (2) perception and training studies must be deepened to better understand how the human ear and brain interpret this type of sound.

\subsection{Stars}
\label{sec:stars}

\subsubsection{Variable stars}
\label{sec:variablestar}

Variable stars have the characteristic that their brightness varies over time, which may be due to characteristics of the star (intrinsic, such as pulsating, eruptive, or cataclysmic variables) or external characteristics (extrinsic, such as eclipsing ones). In the case of this work, examples of intrinsic variable stars were used, of the pulsating type, specifically the Cepheid type (the characteristic of this type of star is that its period is proportional to its luminosity). In addition, a sonification of an extrinsic variable star was provided, of the eclipsing type, which are binary stars whose plane of their orbit coincides with the observation direction, so one star is observed passing in front of the other producing eclipses (the instrument observes a decrease in the amount of received light).

\begin{figure}
    \includegraphics[width=\columnwidth]{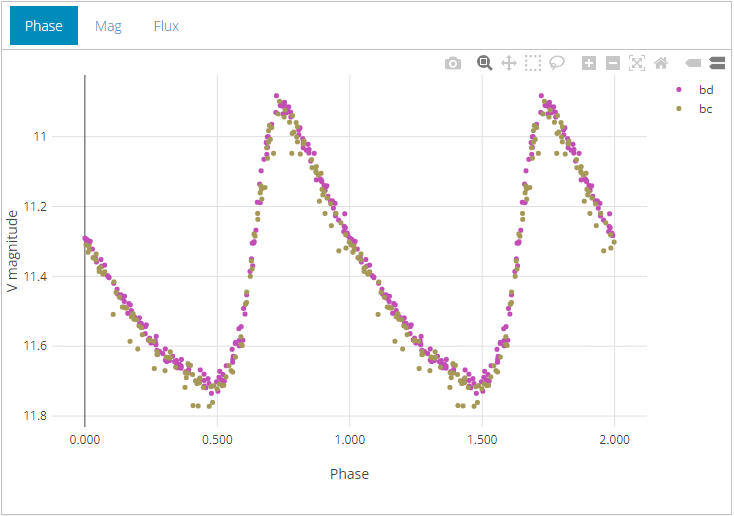}
    \includegraphics[width=\columnwidth]{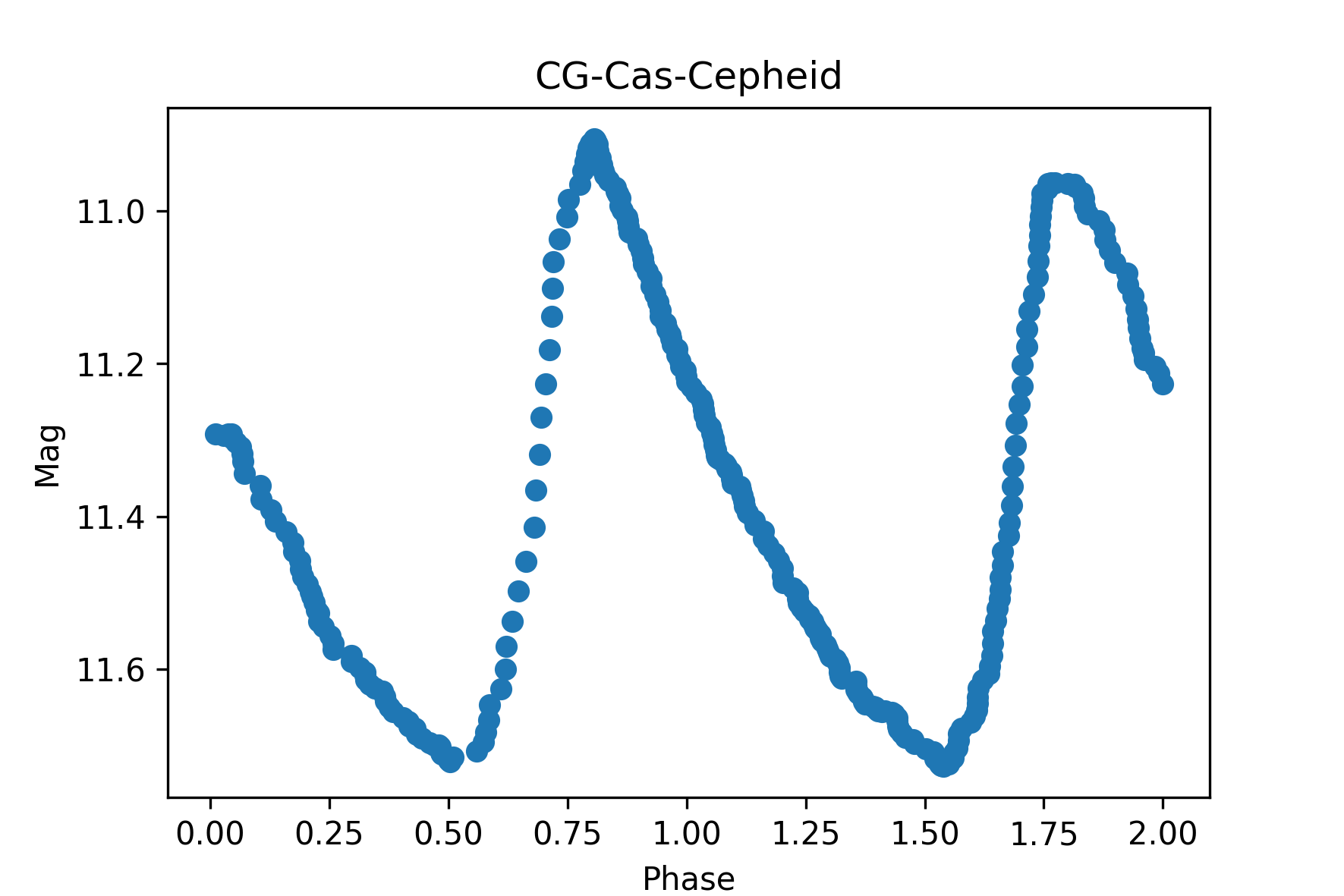}
    \caption{The phase diagram of a Cepheid extracted from the \href{https://asas-sn.osu.edu/variables/753bdd73-38a7-5e43-b6c0-063292c7f28d}{ASAS-SN database} (top). The phase diagram obtained from sonoUno using the same dataset (bottom) (sonification \href{https://www.sonouno.org.ar/wp-content/uploads/sites/9/2023/01/CG-Cas-Cepheid.csv_sound.wav}{here}).}
    \label{fig:asas_cefeidaweb}
\end{figure}

\begin{figure}
    \includegraphics[width=\columnwidth]{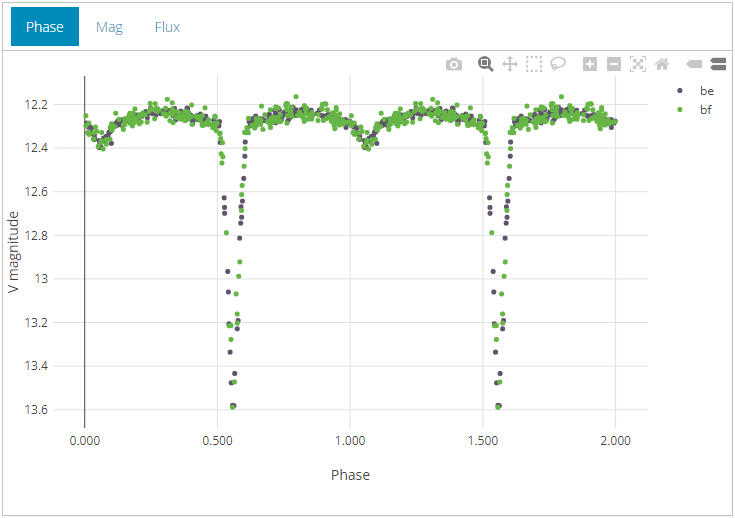}
    \includegraphics[width=\columnwidth]{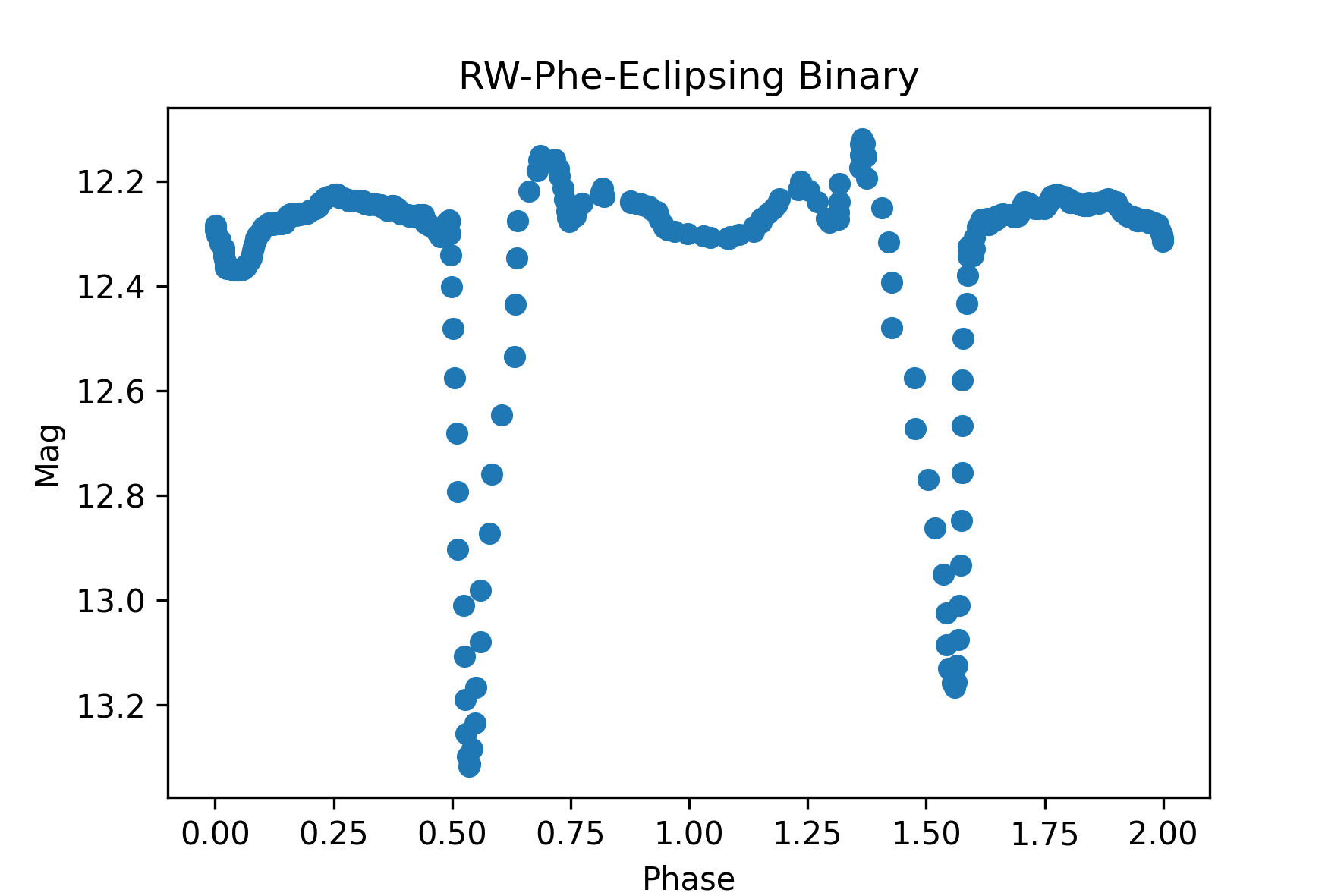}
    \caption{The phase diagram of an Eclipsing Binary extracted from the  \href{https://asas-sn.osu.edu/variables/dfa51488-c6b7-5a03-abd4-df3c28273250}{ASAS-SN database} (top). The phase diagram obtained from sonoUno using the same dataset (bottom) (sonification \href{https://www.sonouno.org.ar/wp-content/uploads/sites/9/2023/01/RW-Phe-Eclipsing-Binary.csv_sound.wav}{here}).}
    \label{fig:asas_eclipsanteweb}
\end{figure}

The ``All-Sky Automated Survey for Supernovae''\footnote{ASAS-SN: https://asas-sn.osu.edu/variables} (ASAS-SN) database contains information on freely accessible variable stars, where a table format file (`csv' extension) with the particular observation data of the star could be downloaded. From this database, the following variable stars were selected to sonify with sonoUno:

\begin{itemize}
    \item Cepheid: ASASSN-V J000059.21+605732.5 / CG Cas (\href{https://asas-sn.osu.edu/variables/753bdd73-38a7-5e43-b6c0-063292c7f28d}{link})
    \item Eclipsing Binary: ASASSN-V J003016.19-462759.5 / RW Phe (\href{https://asas-sn.osu.edu/variables/dfa51488-c6b7-5a03-abd4-df3c28273250}{link})
\end{itemize}

One detail to keep in mind with the data obtained from the variable stars mentioned is that they do not directly have a column with the phase values, because there are different ways to obtain these phase values depending on what you want to plot. In the case of this work, Equation \ref{eq:phase_diagram} was used to calculate the phase, taking into account the epoch based on the `Heliocentric Julian Date' indicated by the same database (`Epoch (HJD)').

\begin{equation}
    \phi = \frac{t - t_{0}}{P}
    \label{eq:phase_diagram}
\end{equation}

As the epoch and period parameters are different for each of the variable stars selected, in Equation \ref{eq:cepheid}, the phase calculation for the Cepheid star can be observed, and in Equation \ref{eq:eclipsing_binary} that corresponding to the Eclipsing Binary star. Finally, using this phase calculation as an independent variable and the magnitude as a dependent variable, the plots presented at the bottom in Figures \ref{fig:asas_cefeidaweb} and \ref{fig:asas_eclipsanteweb} were obtained. In that Figures could be observed the equivalence between the plot on the database and that obtained from sonoUno.

\begin{equation}
    \phi = \frac{t - 2457412,70647}{4,3652815}
	\label{eq:cepheid}
\end{equation}

\begin{equation}
    \phi = \frac{t - 2458053,49761}{5,4134367}
	\label{eq:eclipsing_binary}
\end{equation}

Although the calculus can be carried out externally, to later enter the table into sonoUno and perform the sonification; here a script was presented, it is available in the sonoUno repository with the name `sonify\_bash\_lightcurve.py' (\href{https://github.com/sonoUnoTeam/sonoUno-desktop/blob/0919e706f667962a391f8ef6b64073eb4a99c5c6/sonoUno/sonify_bash_lightcurve.py}{link}) and it used some sonoUno modules. This code allows loading into a variable the value of the time and period referring to the data (currently manually in the code) and then indicating by bash the directory where the file to be sonifyed is located. The script automatically saves the sound in that same folder with the same file name and the appendix `\_sound'. Optionally, the user can indicate to save the plot when executing the code, in which case it will also save it in the same folder and with the same name by adding `\_plot'. The graphics and sounds presented at the bottom in Figures \ref{fig:asas_cefeidaweb} and \ref{fig:asas_eclipsanteweb} were obtained with the script described here.
\\
\\
\textit{Discovery of a variable star}
\\
\\
In 2022, the discovery of the variable star UCAC4 459-09273 using for the first time sonification (and in particular using sonoUno web interface) was made by six blind students in a workshop given by the IES José de Frugoni Pérez in Spain \citep{morales2024}, within the framework of the International Day of Light (news was published on the \href{https://astronomiayeducacion.org/taller-2-de-sonificacion-descubriendo-el-universo/?cn-reloaded=1&cn-reloaded=1}{website} of the Astronomical Association and Educativa de Canarias “Henrietta Swan Leavitt” (AAEC).

These researchers/educators applied a multimodal data analysis strategy, providing the young scientists not only with the data tables to sonify, but also the tactile models of the star field areas where the objects to study were distributed, to identify each star, search for the data, and finally detected the variability if it existed. The details of the discovered variable star are at \href{https://www.aavso.org/vsx/index.php?view=detail.top&oid=2227443}{website} of The International Variable Star Index of the American Association of Variable Stars (AAVSO).

\subsubsection{Spectral Classification}
\label{sec:spectroscopy}

The spectroscopy is a powerful technique in Astrophysics. In general, the spectral classification of stars is performed by visual inspection comparing a spectrum with those corresponding with a standard star. Of course, it is possible to apply algorithms for automatic classification, which provide the results without most of the work done in the past.
However, it is interesting to use the traditional methodology to explore the possibilities for classification using sound.
If the stellar spectrum to classify has a good signal-to-noise ratio, and as sonoUno can compare data from more than one column, the classification is simple, but the technique could be particularly useful if we are in the presence of a bad s/n rate, because it has been proved that in these cases, the sonification can helps to detect features not well defined \citep{wandathesis:2013}.

\begin{figure}
	\includegraphics[width=\columnwidth]{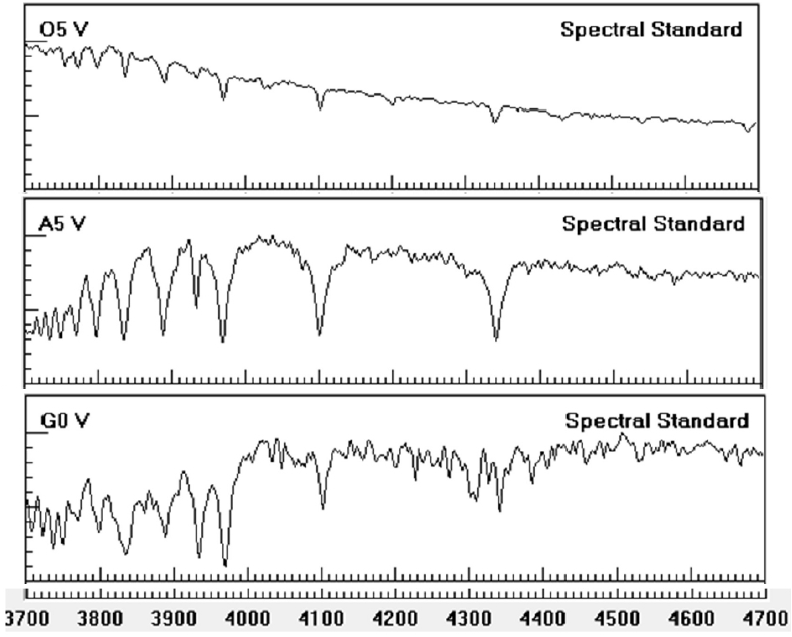}
    \caption{Standard stars spectra for O5V, A5V and G0V types (Credit: Project CLEA).}
    \label{fig:OAG}
\end{figure}

\begin{figure}
    \includegraphics[width=\columnwidth]{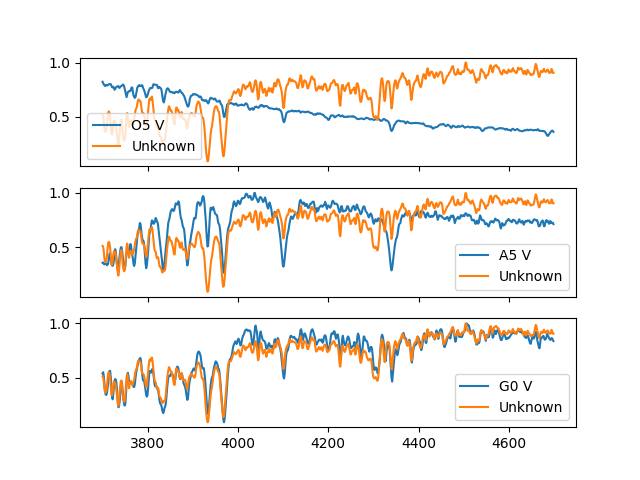}
    \caption{The blue lines sonifyed with 'sine' instrument represent the standard star spectra for O5V, A5V and G0V plotted with sonoUno, comparable with Figure \ref{fig:OAG}. The orange line was from an unknown star to be classified, the video could be accessed \href{https://youtu.be/IHOa7vLaetI}{here}, and the sound of each individual plot: \href{https://www.sonouno.org.ar/wp-content/uploads/sites/9/2024/01/O5-vs-unknown.wav}{O5-unknown}, \href{https://www.sonouno.org.ar/wp-content/uploads/sites/9/2024/01/A5-vs-unknown.wav}{A5-unknown} and \href{https://www.sonouno.org.ar/wp-content/uploads/sites/9/2024/01/G0-vs-unknown.wav}{G0-unknown}.}
    \label{fig:OAG_clasification}
\end{figure}

Some examples of synthetic standard stars spectra are in Figure \ref{fig:OAG}, taken from Project CLEA\footnote{Project CLEA: \url{ http://public.gettysburg.edu/~marschal/clea/}} of Gettysburg College. The plots and the sound, both produced with sonoUno, from data taken from the Atlas of Standard Stars of CLEA, are in Figure \ref{fig:OAG_clasification}. Particularly, the standard data are plotted in blue lines and sonifyed with a `sine' instrument; the sound of each spectrum can be heard in the next links: \href{https://www.sonouno.org.ar/wp-content/uploads/sites/9/2024/01/O5-37004700.wav}{O5V}, \href{https://www.sonouno.org.ar/wp-content/uploads/sites/9/2024/01/A5-37004700.wav}{A5V} and \href{https://www.sonouno.org.ar/wp-content/uploads/sites/9/2024/01/G0-37004700.wav}{G0V}.

The classification of stars' spectra was performed using one sound for the standards (`sine' instrument) and another for the unknown spectra (`flute' instrument). One example of the plot, sound, and video of the classification can be found in Figure \ref{fig:OAG_clasification}. The work was done by trained people in stellar classification.

The bash script for spectroscopy was used here. The algorithm receives the folder path with the unknown spectra and could be configured with three additional parameters: (1) Data file format: as default `txt'; (2) Plot: a flag to save or not the plot (an example of plot in Figure \ref{fig:OAG_clasification}); (3) Display: a flag to show or not the reproduction with the plot and sonification (an example on this \href{https://www.youtube.com/watch?v=IHOa7vLaetI}{video}). Using the display mode OFF, the computer could work hours performing automatically all the images and sonifications to be analyzed later.

\section{Conclusions}

From the results obtained in this work and the methods described to produce the displays (visual and auditory), we could conclude that this technique is reliable and faithfully deploys the data.

In section \ref{sec:galaxies} image sonification was used with a barred spiral and a double nucleus galaxies. Even when the technique is very basic, the increase in light can be perceived through sound. Further, the spectrum obtained from the CSV in sonoUno GUI (Figures \ref{fig:barred_spiral_spectrum1} and \ref{fig:double_nucleus_spectrum1}) is equivalent to that downloaded from the database. The detection of pics, permits the identification of chemical elements in a similar way than using other standard astronomical software, but in a multimodal framework. Besides, the multicolumn sonification was used between these galaxy spectra, with noticeable differences in sound. 

Remarkably, the previous reference of multicolumn sonification was by xSonify, where it is possible to open a multicolumn set of data and perform the sonification of at least three plots, but the graphs are individual, one under the other. This technique had not been replicated by another tool for galaxies, and there has been no evidence of the use of it for the spectral star classification, until now.

As in the standard stellar classification technique, the comparison of known and unknown spectra was shown in a visual mode and as a complement, the data sonification was performed; the links to the audios comparing each standard star in conjunction with the unknown data set are presented in Section \ref{sec:spectroscopy} (Figure \ref{fig:OAG_clasification}). It could be appreciated that the sonification reflects the same as the plot; if there is coincidence, the two sounds fit; and in the case of difference in the spectrum, the two sounds are chaotic. Something that enriches the technique is the possibility of hearing the sound and visualizing the plot at the same time. This novel approach allows us to compare data sets by plot and sound, using the skill of the human ear.

Mainly in this contribution is shown the use of sonification as a complement to the actual visual inspection of the datasets. This enables on one side the comparison between the actual display and sonification, opening the discussion to its reliability; and on the other, start training a new technique.

Although internal tests have been carried out with specialists, it is necessary to perform a broader study, following the example of recent ones developed by \citet{MNRAStucker2022,trayford2023}; the evidence highlights good results after user testing with sonification without a lot of training, given the novelty of the method. More studies should be done to improve understanding of multisensory perception, the use of more than one sense to study the unknown should improve perception and reduce single-sense overload. To be able to measure perception in this sense, sonification training has to be done to level the path between the visual and sonification techniques.

The dual-purpose birth of sonification in astronomy should not be lost. Even when sonification is here to make sense of nature through a new sense, like hearing, it brings accessibility to astronomical discoveries.

\section*{Acknowledgements}

This work was funded by the National Council of Scientific Research of Argentina (CONICET) and has been performed partially under the Project REINFORCE (GA 872859) with the support of the EC Research Innovation Action under the H2020 Programme SwafS-2019-1.
The support from the IBIO-UM and the UTN-FRM is also appreciated, as well as the contribution of the people who participated in testing the software and using it in their research, and the very useful comments by Gonzalo de la Vega and Gonzalo Cayo; the permanent feedback with the interested people contributed significantly to improving this project.
\section*{Data Availability}

sonoUno desktop version is an open-source development and can be used freely: the code,  installation, and user manuals are available on Github: \url{https://github.com/sonoUnoTeam/sonoUno-desktop}. Also, the code of multicolumn sonification is available on GitHub: \url{https://github.com/sonoUnoTeam/sonoUno-multicolumn}.

All the data from different sources used in this work are available in their own databases. The new plots and data sonification made with sonoUno, can be recovered from the sonoUno website (\url{https://reinforce.sonouno.org.ar/}).



\bibliographystyle{apalike}
\bibliography{biblio} 

\end{document}